\documentstyle[12pt,amssymbols]{article}
\def\a{\alpha}
\def\b{\beta}

\def\d{\delta}
\def\l{\lambda}

\def\rmat{R- matrix}
\def\L{L_{a,n}(\lambda)}
\def\n{\nu_a}
\def\h{h_n}

\def\rep{representation }

\def\l{\lambda}
\def\bp{U_q(b_+)}
\def\bm{U_q(b_-)}

\def\RMAT{R=\sum_i A_i\otimes B_i \in U_q(b_+)\otimes U_q(b_-) }

\def\half{{1\over2}}
\def\sl{U_q(\widehat{sl}_2)}
\def\s{U_q(sl_2)}
\def\<{\langle}
\def\>{\rangle}
\newcommand{\bn}{\begin{equation}}
\newcommand{\ed}{\end{equation}}

\newcommand{\Rmt}{ universal $R$--matrix }
\newcommand{\hsp}{\mbox{$\hspace{.5in}$}}

\begin{document}
\rightline{LPTHE-96-11}
\rightline{hep-th/9603105}
\vspace{1.7in}
\centerline{\Large  Quantum Group Representations and
Baxter Equation }
\vskip 1.4cm
\centerline{\large Alexander Antonov$^{1,\,\, 2,}$
\footnote[3]{e-mail address: antonov@lpthe.jussieu.fr and 
antonov@landau.ac.ru}
\,\,\,and \,\,\,
Boris Feigin$^{2,}$
\footnote[4]{e-mail address: feigin@landau.ac.ru}}
\vskip1.2cm
\centerline{ ${}^1$ Laboratoire de Physique Th\'eorique et Hautes
Energies}
\centerline{ Universit\'e Pierre et Marie Curie, Tour 16 1$^{er}$
		\'etage, 4 place Jussieu}
\centerline{75252 Paris cedex 05-France}
\centerline{and}
\centerline{ ${}^2$  Landau Institute for Theoretical Physics }
\centerline{Kosygina 2, GSP-1, 117940 Moscow V-334, Russia }
\vskip1cm
\centerline{\it Dedicated to Memory of Sasha Belov}
\vskip3cm
%%%%%%%%%%%%%%%%%%%%%%%%%%%%%%%%%%%%%%%%%%%%%%%%%%%%%%%%%%%%%%%%%%%

\centerline{\bf Abstract}
\noindent
In this article we propose algebraic universal procedure for deriving 
"fusion rules" and Baxter equation for any integrable model 
with  $U_q(\widehat{sl}_2)$ symmetry
of Quantum 
Inverse Scattering Method. 
Universal Baxter Q- operator is got from the certain
infinite dimensional
representation called q-oscillator one 
of the Universal R- matrix for $U_q(\widehat{sl}_2)$
affine algebra (first proposed by 
V. Bazhanov, S.Lukyanov and A.Zamolodchikov \cite{BLZ1}
for quantum KdV case). 
We also examine the algebraic properties of Q-operator.
\vspace{2in}
\setcounter{section}{-1}

%%%%%%%%%%%%%%%%%%%%%%%%%%%%%%%%%%%%%%%%%%%%%%%%%%%%%%%%%%%%%%%%%%%%%

\section{Introduction}
The problem of diagonalization of an infinite set of mutually commuting 
Integrals of Motion (IM) for some Integrable Quantum Theory is solved
by the Quantum Inverse Scattering Method (QISM) \cite{FST}.
The generating function for IM is trace of monodromy matrix $T(\l)$
depending on some spectral parameter. The powerful method for finding
eigenvalues of 
 $T(\l)$ was proposed by Baxter \cite{Bax}, where he used so called 
$Q(\l)$- operator satisfying Baxter eq.
\bn
T(\l) \, Q(\l)= Q(q \l)+ Q(q^{-1} \l)
\label{TBeq}
\ed
The properties of Q- operator for quantum KdV system were examined
in the paper \cite{BLZ1}.

Another approach for the objects of QISM such as Q- operator or 
monodromy matrix is based on so called \Rmt. An explicit formula 
for \Rmt for quantum affine 
algebras was found in \cite{hor}.
Later we will consider only the case of $\sl$ algebra, though 
the proposed
technique is applicable for general affine quantum algebras.

For $\sl$ algebra the \Rmt lies in the square 
$R\in \sl\otimes \sl$ and satisfies the Yang-Baxter (YB) eq.
$$
R_{12} R_{13} R_{23} = R_{23} R_{13} R_{12}, 
$$
in the $\sl\otimes\sl\otimes\sl$.
We write $R=\sum_i A_i \otimes B_i \in \sl\otimes \sl$ and let  
$$
R_{12}=\sum_i
A_i \otimes B_i \otimes 1 ,\,\,
R_{13}=\sum_i A_i  \otimes 1 \otimes B_i ,\,\,
R_{23}=\sum_i  1 \otimes A_i \otimes B_i . 
$$
Monodromy matrix can be got from the \Rmt
$R\in \sl\otimes \sl$ using matrix $n\times n$ representation 
with a spectral parameter
for the first $\sl$ algebra and some other representation
(depending on a integrable model in consideration)
for second $\sl$ algebra in the square  $\sl\otimes\sl$.
First, it was noticed by L.Faddeev in \cite{SB}.
In the same way, it is possible to get Q- operator
from the \Rmt $R\in \sl\otimes \sl$. For this we choose certain
{\it infinite dimensional} representation for the first $\sl$ 
algebra.

Using Q- operator or monodromy matrix presentation from the \Rmt
one clarifies the "fusion rules" and Baxter eq. The origin of 
these eqs.
is in certain $\sl$ representation square.
Namely, we have ($M_1$ and $M_2$) two QISM objects (monodromy matrix 
or Q- operators)
got from \Rmt using two certain representations ($W_1$ and $W_2$)
 for the first $\sl$ algebra in 
$R\in \sl\otimes \sl$.
Then, there is one-to-one correspondence between "fusion" like 
relation
for $\hbox{tr}_{W_1} M_1$ and  $\hbox{tr}_{W_2} M_2$
and the structure of the representation square $W_1\otimes W_2$.
We will explain how using the existence of a subrepresentation 
in square
 $W_1\otimes W_2$ one can derive "fusion rules" and Baxter eq.

According to the methods of \Rmt for monodromy matrix or Q- operators
we specify representation only of the {\it first} $\sl$ algebra
for  $R\in \sl \otimes  \sl$, and representation of the second one
(depending on a concrete model in hand) is not fixed.
So, we propose {\it universal} technique for deriving Baxter eq.
 as well as for algebraic understanding of Baxter Q- operator,
which is very important in the Algebraic Bethe Ansatz.

The text is organized as follows.
In {\bf Section 1} we remind some key moments of QISM and interpret
monodromy matrix via \Rmt representations. 
We introduce algebraic method for constructing "fusion" like relations
and give some examples for finite dimensional $\sl$ representations in 
{\bf Section 2}.
We define {\it infinite dimensional} representation for Baxter 
Q- operator from \Rmt
in {\bf Section 3}. There we derive algebraically Baxter eq. 
and some other
 "fusion" like eqs. for Q- operator. The discussion and conclusion
are presented in {\bf Section 4}.

%%%%%%%%%%%%%%%%%%%%%%%%%%%%%%%%%%%%%%%%%%%%%%%%%%%%%%%%%%%%%%%%%%%%
\vspace{0.2in}
{\bf Acknowledgments.}
A.A. is thankful to L.Baulieu and
Laboratoire de Physique Th\'eorique et Hautes \'Energies,
Paris, France for the hospitality.
The authors would like to thank A.Zamolodchikov for helpful discussions
and for sharing with them the results of the paper \cite{BLZ1} prior 
to its publication.
A.A. is grateful to  A.Belavin, L.Chekhov, B.Enriquez, L.Faddeev, R.Kashaev,
M.Lashkevich, V.Rubtsov, F.Smirnov,
A.Volkov, A.Zabrodin
for illuminating discussions.

%%%%%%%%%%%%%%%%%%%%%%%%%%%%%%%%%%%%%%%%%%%%%%%%%%%%%%%%%%%%%%%%%%

\section{ Origin of QISM Ingredients from Universal R-matrix }
In this section we use the notations from \cite{SB}.
The quantum L-operator is the main object of the QISM.
$\L$ is a matrix in an auxiliary space $\n$, the matrix 
elements of $\L$
are operators in a Hilbert space $\h$, 
$n=1, \cdots , N $
associated with the site of the lattice, depending on the 
spectral parameter
$\l$.
The operators in different sites commute.
In this way, the operator $\L$ is the operator in the square 
 $\n\otimes\h$.

The fundamental commutation relations for the matrix 
elements of $\L$
is the YB eq.:
\begin{equation}
R_{a_1,a_2}(\lambda / \mu) L_{a_1,n}(\lambda) L_{a_2,n}(\mu)=
L_{a_2,n}(\mu) L_{a_1,n}(\lambda)R_{a_1,a_2}(\lambda / \mu). 
\label{fcr}
\end{equation}
This is an equation in $\nu_{a_1} \otimes \nu_{a_2} \otimes h_n$. 
The indices $a_1$ and
$a_2$ and the variables $\lambda$ and $\mu$ are associated with
the auxiliary spaces $\nu_1$ and $\nu_2$, respectively.
The matrix $R_{a_1,a_2}$ is one in the space 
$\nu_{a_1} \otimes \nu_{a_2} $.

Further, we consider only the special form of $R_{a_1,a_2}$- matrix, 
so called
trigonometric R-matrix, $4 \times 4$. A lot of interesting 
models on a lattice
(XXZ spin model, the lattice Sine- Gordon system, 
Volterra system etc.)
and in continuum ( quantum mKdV system \cite{BLZ2}) 
are described by the 
trigonometric R-matrix.   

Another important object of the QISM is the monodromy matrix
\begin{equation}
M_a(\lambda) = L_{a,N}(\lambda) L_{a,N-1}(\lambda) \cdots  L_{a,1}(\lambda).
\label{tt}
\end{equation}
The monodromy matrix satisfies commutation relations identical 
with the  YB ones
for L- operators (\ref{fcr}). It follows that the traces 
$T_a(\l)=\hbox {tr}_{\n} M_a (\l)$ over the auxiliary spaces
of the monodromy 
matrix with the different parameters commute 
$[T_{a_1}(\l), T_{a_2}(\mu)]=0$.
That is why the trace of the monodromy matrix is the 
generating function for the
integrals of motion.

It is useful to consider so called fundamental 
L- operator \cite{FTT},
$L_{n_1,n_2}(\l)$, i.e. the operator in $h_{n_1}\otimes h_{n_2}$. 
In other words the auxiliary space coincides
with the quantum one. The YB eq. for the fundamental 
L-operator is following
\begin{equation}
L_{a,n_1}(\l) L_{a,n_2}(\mu) L_{n_1,n_2}(\mu/\l)=
L_{n_1,n_2}(\mu/\l) L_{a,n_2}(\mu) L_{a,n_1}(\l), 
\label{fyb}
\end{equation}
as the eq. in $\nu_a\otimes h_{n_1}\otimes h_{n_2}$.
The monodromy matrix for the fundamental L-operator 
gives the set of local
integrals of motion.
%%%%%%%%%%%%%%%%%%%%%%%%%%%%%%%%%%%%%%%%%%%%%%%%%%%%%%%%%%%%%

It is possible to describe all the objects of QISM 
(trigonometric \rmat,
L- operator and monodromy matrix as well as 
fundamental L- operator)
with the only algebraic object- \Rmt  \cite{SB}.

We will briefly remind some facts about it \cite{hor}. 
Below, we consider
the simple case of the affine quantum $\sl$ algebra 
with Cartan matrix 
 $A=(a_{ij})$, $i,j=0,1$,
$
A = \left( \begin{array}{cc}
2 & -2 \\
-2 &  2
\end{array}
\right) $.
%%%%%%%%%%%%%%%%%%%%%%%%%%%%%%%%%%%%%%%%%%%%%%%%%%%%%%%
This is an associative
algebra with generators 
$e_{\pm \alpha_{i}},\; k_{\alpha_{i}}^{\pm 1},\;
 (i=0,1)$, and the defining relations
$$
[k_{\alpha_{i}}^{\pm 1}, k_{\alpha_{j}}^{\pm 1}]=0,
\hspace{.5in}
k_{\alpha_{i}}e_{\pm \alpha_{j}}= q^{\pm (\alpha_{i},\alpha_{j})}
e_{\pm\alpha_{j}} k_{\alpha_{i}},
$$
$$
[e_{\alpha_{i}}, e_{-\alpha_{j}}] = \delta_{ij}\frac{k_{\alpha_{i}} -
k_{\alpha_{i}}^{-1}}{q-q^{-1}},
$$
$$
(ad_{q'}e_{\pm\alpha_{i}})^{1-a_{ij}} e_{\pm\alpha_{j}}=0 
\hspace{.5in}
\hbox {for}\;\,\,\, i\neq j, \:q'=q,q^{-1},
$$
where $(ad_{q}e_{\alpha})e_{\beta}$ is a q-commutator:
$$
(ad_{q}e_{\alpha})e_{\beta}\equiv
{[e_{\alpha},e_{\beta}]}_{q}=e_{\alpha}e_{\beta}-q^{(\alpha,\beta )}
e_{\beta}e_{\alpha}
$$
and  $(\alpha ,\beta )$ is a scalar  product of the roots $\alpha$
and $\beta$: $(\alpha_{i},\alpha_{j})=a_{ij}$.
The condition $ 1=k_{\a_0} k_{\a_1}$ is also imposed.

We define a comultiplication in $\sl$ by the formulas
$$
\Delta (k_{\alpha_{i}})=k_{\alpha_{i}}\otimes k_{\alpha_{i}},
$$
\bn
\Delta (e_{\alpha_{i}}) = e_{\alpha_{i}}\otimes
k_{\alpha_{i}} + 1 \otimes e_{\alpha_{i}}
\label{comult}
\ed
$$
\Delta (e_{-\alpha_{i}}) = e_{-\alpha_{i}}\otimes 1+
k^{-1}_{\alpha_{i}}\otimes e_{-\alpha_{i}},
$$
and the twisted comultiplication
$$
\Delta '(k_{\alpha_{i}})=k_{\alpha_{i}}\otimes k_{\alpha_{i}},\;
$$
$$
\Delta' (e_{\alpha_{i}}) =k_{\alpha_{i}}\otimes e_{\alpha_{i}} + 
e_{\alpha_{i}}\otimes 1,
\,\,\,\,
\Delta ' (e_{-\alpha_{i}}) = 1\otimes e_{-\alpha_{i}}+
 e_{-\alpha_{i}}\otimes  k^{-1}_{\alpha_{i}},
$$

By the definition, the \Rmt is an object $R$ in $\sl\otimes\sl$ such that
$\Delta '(g) R = R \Delta (g)$, for any $g \in \sl$ and 
\begin{equation}
R_{12} R_{13} R_{23} = R_{23} R_{13} R_{12}, 
\label{uyb}
\end{equation}
which is the universal form of the YB eq. in the $\sl\otimes\sl\otimes\sl$.
We write $R=\sum_i A_i \otimes B_i \in \sl\otimes \sl$ and let  
$$
R_{12}=\sum_i
A_i \otimes B_i \otimes 1 ,\,\,
R_{13}=\sum_i A_i  \otimes 1 \otimes B_i ,\,\,
R_{23}=\sum_i  1 \otimes A_i \otimes B_i . 
$$
In {\bf Appendix A} we present the explicit form of the \Rmt for $\sl$
following \cite {hor}.
We know (\ref{urm}) that \Rmt belongs to the square of $\bp\otimes\bm$,
where $b_\pm$ are positive (negative) borel subalgebras of $\sl$,
generated by 
$e_{ \alpha_{i}},\; k_{\alpha_{i}}^{\pm 1}$
and
$e_{ -\alpha_{i}},\; k_{\alpha_{i}}^{\pm 1}
,\; (i=0,1)$, respectively.

One can present QISM L-operators from the \Rmt using the certain
representation of $\sl$. For example, the trigonometric $4\times 4$ \rmat
is got from the \Rmt representing $\sl$ in terms of matrix $2\times 2$
with the spectral parameter.

It is possible to get the $\L$ operator from the \Rmt 
$R\in \bp\otimes\bm$ representing 
the first algebra $\bp$
in matrixes with the spectral parameter and
the second one $\bm$
in quantum (infinite dimensional) space.
Substituting the comultiplied $\Delta^{(N-1)}\sl$ quantum space 
generators in the
$\bm$ part of the \Rmt we get the monodromy matrix of N-sites (\ref{tt}).

Now we illustrate two different forms of the YB eq. (\ref{fcr}), (\ref{fyb})
from the point of view of the universal one (\ref{uyb}). Eq. (\ref{uyb})
is one for the algebraic elements in $\sl\otimes\sl\otimes\sl$.
To get the eq.  (\ref{fcr}) we represent the first and the second algebras
$\sl$ in finite dimensional (with respect to $\s$) representation, say 
matrix with the spectral parameter and the third in quantum space $\h$.
The eq. (\ref{fyb}) is the representation of the eq.  (\ref{uyb})
with
the first algebra  in matrixes with the spectral parameter and 
the second and the third algebras 
in the quantum spaces $h_{n_1}$  and $h_{n_2}$.
Below we will consider the ingredients of QISM got from the \Rmt
$\RMAT$ for certain representation of the first $\sl$ algebra 
and for {\it any}
representation of the second $\sl$ one (for {\it any} 
integrable model
with $\sl$ symmetry).

%%%%%%%%%%%%%%%%%%%%%%%%%%%%%%%%%%%%%%%%%%%%%%%%%%%%%%%%%%%%%%%%%%%%%%%%

\section{ "Fusion Rules" in Algebraic Presentation}

The treatment of the origin of the QISM L-operators as the certain 
representation of the \Rmt gives possibility for an algebraic 
comprehension of the "fusion rules" and the Baxter eq. 
For these relations
we use only the first algebraic elements $A_i$ of the \Rmt $\RMAT$,
so nothing depends on a representation of the second $\sl$ algebra
(algebraic elements $B_i$)
and, consequently, on the concrete model of the QISM.

Consider the evaluation representation for the $\sl$ in terms of 
$\s\otimes \Bbb {C}[\l,\l^{-1}]$
\begin{equation}
e_{\pm \alpha_{1}}= e_{\pm \alpha},
\hspace{.5in}
e_{\pm \alpha_{0}}= \l^\pm e_{\mp \alpha},
\hspace{.5in}
k_{\alpha_{1}}= k,
\label{ev}
\end{equation}
where $\s$ algebra has generators $ e_{\pm \alpha} $ and $k^{\pm 1}$
with the ordinary commutation relations
$$
k e_{\pm \alpha}= q^{\pm 2}
e_{\pm\alpha} k,
\hspace{.5in}
[e_{\alpha}, e_{-\alpha}] = \frac{k - k^{-1}}{q-q^{-1}},
$$

Let $V_j$ be the irreducible $2j+1$ -dimensional representation of $\s$,
$j=0, \half, 1, \cdots$.
One can define the representation $ V_j(\l)$ of the affine $\sl$ algebra 
from (\ref{ev})
and  the  representation $V_j$ of the simple $\s$ algebra. 

Let us denote by $M_j(\l)$ some L-operator or monodromy matrix got
from \Rmt $\RMAT$ by following representation
\begin{itemize} 
\item for the first algebra $\bp$ we take the evaluation representation  $ V_j(\l)$ 
of $\s$ spin $ j $ ;
\item the second algebra $\bm$ of the \Rmt is represented in some
quantum space, depending on a physical model in consideration.
\end{itemize} 
Let $T_j(\l)= \hbox{tr}_{ V_j(\l) } M_j(\l)$ be the trace of monodromy matrix.
In fact, 
\bn
[T_j(\l),T_{j'}(\l')]=0 ,
\label{comm}
\ed
that is why  $T_j(\l)$  for any spin  $j$  are generating functions
for integrals of motion. 
There is an algebraic dependence between traces of monodromy matrixes
with different spin  $j$, so called "fusion rules"
\bn
T_j (q \lambda)\; T_j (q^{-1}\lambda) =
1+ T_{j-{1\over 2}}(\lambda) \;T_{j+{1\over 2}}(\lambda)
\label{fr}
\ed

{\it  Remark} 
\hspace{.3in}
The fusion procedure (\ref{fr}) differs from the ordinary notations by 
$ T(\l^2)= T'(\l)$. The origin of this is in different evaluation 
representations. For example, if we write the representation  as
$$
e_{\pm \alpha_{1}}= \l^{\pm 1} e_{\pm \alpha},
\hspace{.5in}
e_{\pm \alpha_{0}}= \l^{\pm 1} e_{\mp \alpha},
\hspace{.5in}
k_{\alpha_{1}}= k,
$$
we get the ordinary form of fusion rules. Below for the reasons
of algebraic simplicity we will consider the representation (\ref{ev}).

Let us illustrate the fusion rules  (\ref{fr}) in algebraic way.
We will claim that eq. (\ref{fr}) can be derived from a consideration
of square of $\sl$ representations.

Indeed, there exists the trivial representation  \,{\bf 1} in the square
$$V=V_j(\l q^{-1}) \otimes V_j(\l q)$$. 
The factor $V/$ {\bf 1}
is isomorphic to 
$$ 
V_{j-{1\over 2}}(\lambda)\otimes  V_{j+{1\over 2}}(\lambda).
$$ 
Consider now two \Rmt
$R_{1,2}=\sum_i A_i^{(1)}\otimes B_i$ and
$R_{1',2}=\sum_n A_n^{(1')}\otimes B_n$,
acting in
$V_j (\l q^{-1}) \otimes h$ and
$V_j (\l q) \otimes h$, respectively.
The Hilbert space $h$ depending on a model in consideration is not fixed.

By the definition,
\bn 
T_j (q^{-1}\lambda) \;T_j (q \lambda) =
\hbox {tr}_{V_j(\l q^{-1}) \otimes V_j(\l q) }
\sum_{i,n} A_i^{(1)}\otimes  A_n^{(1')}\otimes B_i B_n
\label{sq}
\ed
To take the trace over $V=V_j(\l q^{-1}) \otimes V_j(\l q)$
one needs two steps:
\begin{itemize} 
\item to calculate the trace over the trivial subrepresentation \,{\bf 1} of the representation   $V$
(it gives us 1 in fusion rules (\ref{fr}));
\item take the trace over the factor space 
$V/ \hbox{{\bf 1}}
\cong V_{j-{1\over 2}}(\lambda)\otimes  V_{j+{1\over 2}}(\lambda)$
(we get the term 
$ T_{j-{1\over 2}}(\lambda) \,T_{j+{1\over 2}}(\lambda)$ )
\end{itemize} 
In this way we derived the fusion procedure (\ref{fr})
extracting  subrepresentation from  square of certain $\sl$ representations.
One can get other fusion rules from the following fact \cite{ChP}:

if $b/a= q^{\pm (2j+2k-4r+2)}$ for the square
$V=V_j (a)\otimes V_k (b)$,
\,\,\,  
($j\geq k$) of $\sl$ 
representations $(0 < r \leq \hbox{min} (j,k))$,
then V has a unique proper subrepresentation W:
\begin{enumerate}
\item if  $b/a= q^{ (2j+2k-4r+2)}$ , we have
$$
W\cong V_{j-r} (q^{-2r}a)\otimes V_{k-r} (q^{2r}b),
$$
$$
V/W \cong  V_{r-{1\over 2}} (q^{2j-2r+1}a)
\otimes V_{j+k-r+{1\over 2}} (q^{-(2j-2r+1)}b);
$$
\item if  $b/a= q^{-(2j+2k-4r+2)}$ , we have
$$
W \cong  V_{r-{1\over 2}} (q^{-(2j-2r+1)}a)
\otimes V_{j+k-r+{1\over 2}} (q^{2j-2r+1}b),
$$
$$
V/W \cong V_{j-r} (q^{2r}a)\otimes V_{k-r} (q^{-2r}b).
$$
\end{enumerate}
For example, repeating procedure (\ref{sq}) for the first case of $b/a$
one gets the following fusion rules
\bn
T_j (a) \, T_k (b)=
T_{j-r} (q^{-2r}a) \, T_{k-r} (q^{2r}b)+
T_{r-{1\over 2}} (q^{2j-2r+1}a) \,
T_{j+k-r+{1\over 2}} (q^{-(2j-2r+1)}b).
\label{nfr}
\ed

This method is consistent with the commutativity
(\ref{comm}). 
Namely, let us put $j=k$ in eq. (\ref {nfr}).
As we know the consideration of the square $V_j(a)\otimes V_j(b)$
leads us to eq.  (\ref {nfr}) for  $j=k$. 
Another possibility to get rules  (\ref {nfr}) is to extract subrepresentation
from twisted square  $V_j(b)\otimes V_j(a)$.
The difference with the ordinary square  $V_j(a)\otimes V_j(b)$
is in interchanging of sub- and factor-representations.
So, we get the same eq. (\ref {nfr}) from the twisted square
for commuting
$T_j (a)$ and $T_j (b)$.

We see now that the procedure for deriving the fusion rules
is identical to the extracting of some subrepresentation from the square
of certain $\sl$ representations. Of course, it is not limited by the finite dimensional
\, representations $V_j$. In the next section we will see how to construct Baxter operator
$Q(\l)$ from the \Rmt using {\it infinite} dimensional $\bp$ representations.

%%%%%%%%%%%%%%%%%%%%%%%%%%%%%%%%%%%%%%%%%%%%%%%%%%%%%%%%%%%%%%%%%%%%%%%%%%%%
\section{Infinite Dimensional Representations and Baxter Equation}

An important object in QISM is so called Baxter operator $Q(\l)$ \cite{Bax}.
It can be defined in statistical and integrable models via Baxter eq.
\bn
Q(\l)\, T_\half(\l)= Q(q^2 \l)+ Q(q^{-2} \l)
\label{Beq}
\ed
(The difference with the standard Baxter eq. (\ref{TBeq}) is explained
in {\it Remark} in {\bf Section 2}).

The Baxter eq. plays an important role in Bethe Ansatz formalism,
$Q(\l)$ has a simple eigenvalues on Bethe vectors. So, the algebraic 
interpretation of Baxter eq. is useful for getting Bethe vectors in 
purely algebraic terms (i.e. in terms of $\sl$ representation  space).

Using method of {\bf Section 2} and eq. (\ref{sq}) one can get $Q(\l)$ operator
from  $\RMAT$ by following representations   of the first algebra $\bp$ 
 
\bn
e_{\a_0} e_{\a_1}- q^2 e_{\a_1} e_{\a_0}= \frac {\l}{q^{-2}-1}
\hspace{.5in}
(\hbox{the representation } V_+(\l))
\label{rep1}
\ed  

and
 
\bn
e_{\a_1} e_{\a_0}- q^2 e_{\a_0} e_{\a_1}= \frac {\l}{q^{-2}-1},
\hspace{.5in}
(\hbox{the  representation } V_-(\l)),
\label{rep2}
\ed
we have
$$
Q_+(\l)=\hbox{tr}_{ V_+(\l) } R
\hspace{.5in}
Q_-(\l)=\hbox{tr}_{ V_-(\l) } R
$$
for {\it any} representation  of $\bm$ (for {\it any} integrable model)
\footnotemark
\footnotetext[1]
{We choose the infinite dimensional \rep space $V_\pm (\l)$ such that the trace
$Q_\pm(\l)=\hbox{tr}_{ V_\pm(\l) } R$  exists.}.
Such Q- operators satisfies Baxter eq. (\ref{Beq})
and commute with each other and $T_j(\l)$. The representations 
(\ref{rep1}-\ref{rep2}) called q-oscillator algebras
were first used in \cite{BLZ1} for
constructing of Q-operators in KdV system.
 
We illustrate now Baxter eq. (\ref{Beq}) in terms of procedure  (\ref{sq})
for $Q_+$- operator (analogously, for $Q_-$- operator.)
Let two universal R- matrixes 
$$
R_{1,2}=\sum_i A_i^{(1)}\otimes B_i \hbox { and }
R_{1',2}=\sum_n A_n^{(1')}\otimes B_n,
$$
act in spaces
$V_+(\l) \otimes h$ and
$V_\half (\l) \otimes h$, 
respectively,
(the Hilbert space $h$ is arbitrary). For traces
$$
\hbox{tr}_{V_+(\l)}R_{1,2} \hbox { and }
\hbox{tr}_{V_\half (\l)}R_{1',2}
$$
one has
\bn 
Q_+(\lambda)\,\,
T_\half (\lambda)=
\hbox {tr}_{ V_+(\l)\otimes V_\half (\l) }
\sum_{i,n} A_i^{(1)}\otimes  A_n^{(1')}\otimes B_i B_n
\label{sqb}
\ed
For calculating the trace over the square $ V_+(\l)\otimes V_\half (\l)$
we should examine it for an existence of some subrepresentation.

Indeed, it is possible to show (see {\bf Appendix B}),
that the square  $ V_+(\l)\otimes V_\half (\l)$ has the subrepresentation
$ V_+(q^{-2} \l)$. The factor representation 
 $ V_+(\l)\otimes V_\half (\l) / V_+(q^{-2}\l)$  is isomorphic
to $V_+(q^2 \l)$
\bn
V_+(\l)\otimes V_\half (\l) / V_+(q^{-2}\l)\cong
V_+(q^2 \l).
\label{sub}
\ed
To prove Baxter eq. from eqs. (\ref{sqb}) and (\ref{sub})
one needs two usual steps
\begin{itemize}
\item to take trace over the subrepresentation $ V_+(q^{-2} \l)$
(to get the term $ Q_+(q^{-2} \l)$);
\item  to take trace over the factor-presentation $ V_+(q^{2} \l)$
(to get the term $ Q_+(q^{2} \l)$).
\end{itemize}

Baxter operator for the representations (\ref{rep1}) and (\ref{rep2}) 
was derived
analytically  by V.Bazhanov, S.Lukyanov and A.Zamolodchikov 
\cite{BLZ1} for Quantum KdV Model. To get this 
Q-operator from \Rmt $\RMAT$   
one needs to represent $\bm$ algebra via vertex operators
$$
e_{\a_1}= \int \,du\,:e^{ 2\varphi (u)}:
\hspace{.5in}
e_{\a_0}= \int \,du \,:e^{ -2\varphi (u)}:,
$$
where $\varphi (u)$ is the standard bosonic field \cite{BLZ1} , \cite{BLZ2}.

The result of A.Volokov \cite{VQ}  (the coincidence of Q- operator and
trace of fundamental L- operator $L_{n_1,n_2}(\l)$ 
(\ref{fyb}) for Volterra model
\cite{VF}) can be interpreted in terms of representations of \Rmt.
The quantum space (the Hilbert space $h$) generators for Volterra L- operator
satisfies the relations  (\ref{rep1}). So, Q- operator for Volterra model
can be got from $\RMAT$ by using {\it the same} representations  for 
$\bp$ and $\bm$, and thus,
$$
Q(\l)=\hbox {tr}_{h_{n_1}} L_{n_1,n_2}(\l).
$$

Now the natural question arises:
are there any subrepresentation in infinite dimensional squares
$V_\pm(a)\otimes V_\pm(b)$ or  
$V_\pm(a)\otimes V_\mp(b)$ 
with certain $a$  and $b$?
In other words, are there any relations for squares of Baxter operators?
The answer is positive.

Consider the square
$V_+(q^{-1}\l)\otimes V_-(q\l)$.
It is possible to prove, that there exists the set of trivial
subrepresentations {\bf 1}. The factor over it gives
$$
V_+(q^{-1}\l)\otimes V_-(q\l)/\hbox{{\bf 1}}\cong
V_+(q \l)\otimes V_-(q^{-1}\l).
$$ 
Using our technique we derive the following eq.
\bn
Q_+(q^{-1}\l) \, Q_-(q\l)= \hbox{const}+
Q_+(q \l) \, Q_-(q^{-1}\l).
\label{norm}
\ed
Here the const depends on a normalization of representation spaces $V_\pm(\l)$.
Eq. (\ref{norm}) was first obtained for KdV system in \cite{BLZ1}.

From the eqs. (\ref{Beq}) and  (\ref{norm}) we can derive another dependence
between $T_\half$  and Q- operators
\bn
\hbox{const} \cdot
T_\half(\l)=
Q_+(q^{-2}\l) \, Q_-(q^2\l)-
Q_+(q^2 \l) \, Q_-(q^{-2}\l).
\label{spin}
\ed
It is possible also to get this eq. considering representation  square
$V_+(q^{-2}\l)\otimes V_-(q^2 \l)$. Indeed, there exists subrepresentation,
isomorphic to $V_\half(\l)$, and after factorizing we have
$$
V_+(q^{-2}\l)\otimes V_-(q^2\l)/V_\half(\l)
\cong V_+(q^{2}\l)\otimes V_-(q^{-2}\l).
$$

Eq. (\ref{spin}) is the special case of the general formula for $T_j(\l)$
(\cite {BLZ1} for KdV system)
$$
\hbox{const} \cdot
T_j(\l)=
Q_+(q^{-(2j+1)}\l) \, Q_-(q^{(2j+1)}\l)-
Q_+(q^{(2j+1)} \l) \, Q_-(q^{-(2j+1)}\l),
$$
which can be proved fac\-tor\-iz\-ing the square
$V_+(q^{-(2j+1)}\l)\otimes V_-(q^{(2j+1)}\l)$
over the subrepresentation $V_j(\l)$
$$
V_+(q^{-(2j+1)}\l)\otimes V_-(q^{(2j+1)}\l)/V_j(\l)
\cong 
V_+(q^{(2j+1)}\l)\otimes V_-(q^{-(2j+1)}\l).
$$

It is not clear now, are there other
relations connecting $Q_\pm$- operators?
It seems, that considering squares $V_+(a)\otimes V_+ (b)$
for proper $a$  and $b$  one can get sub- or factor representations 
differing from $V_j(\l)$  and  $V_\pm(\l)$. That is why, one can treat
$V_\pm$  (or Baxter operators)
as basic generating representations  (operators).

%%%%%%%%%%%%%%%%%%%%%%%%%%%%%%%%%%%%%%%%%%%%%%%%%%%%%%%%%%%%%%%%%%%%%%%%
\section{Discussion and Concluding Remarks}
In this article we proposed the universal procedure for constructing 
"fusion" like eqs. for quantum monodromy matrix and Baxter Q- operator
for some Integrable system.

We got all the ingredients of QISM from \Rmt. The Q- operator
was presented as certain infinite dimensional $\sl$ representation
(\ref{rep1}-\ref{rep2}) for the first algebra of
$\RMAT$ (the second algebra representation in the square is arbitrary).
This gave us opportunity to treat Baxter eq. in invariant terms,
i.e. we derived it considering certain representation 
squares.
The important point of the method is that we did not specify
the model in hand, all the equations are valid for
any Integrable Systems with the $\sl$ symmetry.

The algebraic origin of QISM objects from \Rmt
gives possibility to:

\begin{itemize}
\item treat $V_\pm(\l)$ representations ($\hbox{Q}_\pm$- operators)
as basic representations (operators), 
and generate new infinite dimensional subrepresentation
(corresponding to new objects of QISM) from the square
$V_\pm(a)\otimes V_\pm(b)$;
\item find Bethe eigenvectors in invariant terms
(i.e. in terms of certain quantum representation
 space).
\end{itemize}

It seems that there are several directions for continuation:
\begin{itemize}
\item the examination of representation
 squares 
$V_\pm(a)\otimes V_\pm(b)$ for existing of subrepresentations;
\item the examination of integrable models with $V_\pm(\l)$ symmetry
(i.e. got from \Rmt by representation
 of second $\bm$ algebra
in $V_\pm$ spaces).
In this case we have coincidence of fundamental R- matrix
and Q- operator. The special case of these system is Volterra one
(representation
 $V_+$ degenerates in commuting 
$e_{\a_1}$ and $e_{\a_0}$ generators, 
$e_{\a_1} e_{\a_0}= e_{\a_0} e_{\a_1}= \frac{\l}{(q-q^{-1})^2}$ ).
\item the generalization of the results to other affine algebras.
\end{itemize}

%%%%%%%%%%%%%%%%%%%%%%%%%%%%%%%%%%%%%%%%%%%%%%%%%%%%%%%%%%%%%%%%%%%%%%%%%%%%

\appendix
\vspace {3.5in}
{\center\Large{\bf Appendix}}

\section { Universal R- matrix for $\sl$ algebra}
In this Appendix we present the explicit form of the \Rmt for $\sl$
following \cite {hor}.

Along with the commutation relations for $\sl$ we need 
an antiinvolution  $(^{*})$ ,
defined as
$(k_{\alpha_{i}})^{*}=k_{\alpha_{i}}^{-1},\;$
$(e_{\pm\alpha_{i}})^{*}=e_{\mp\alpha_{i}},\;$
$(q)^{*}=q^{-1}$.

We use also the following standard notations
$$
\hbox{exp}_{q}(x) := 1 + x + \frac{x^{2}}{(2)_{q}!} + \ldots +
\frac{x^{n}}{(n)_{q}!} + \ldots = \sum_{n\geq 0} \frac{x^{n}}{(n)_{q}},$$

$$(a)_{q}:=\frac{q^{a}-1}{q-1},\,\,\,\,\,\,\,\,\,\,\,\,
[a]_{q}:=\frac{q^{a}-q^{-a}}{q-q^{-1}},\,\,\,\,\,\,\,\,\,\,\,\,
q_{\alpha}:=q^{-(\alpha,\alpha)}$$

We define the Cartan-Weyl generators of the $\sl$.
Let $\alpha\equiv \alpha_1$ and $\beta\equiv \alpha_0=\delta -\alpha$ 
are simple roots for the affine
algebra $\widehat{sl_{2}}$ then $\delta=\alpha+\beta$ is a minimal
imaginary root.  We fix the following normal ordering in the system of the
positive roots: \
$$
\alpha,\;\alpha+\delta, \alpha+2\delta\ldots ,
\delta,\; 2\delta, \ldots , \ldots ,
\b +2\d ,\b +\d ,\b \ .
$$
We put
$$
e'_{\delta}=e_{\delta}=
[e_{\alpha},e_{\b}]_{q} \ ,
$$
$$
e_{\alpha+l\delta}=(-1)^{l}
\left( [(\alpha ,\alpha )]_{q}\right)^{-l}
(ad\ {e'}_{\delta})^{l}e_{\alpha} \ ,
$$
$$
e_{\b+l\delta}=
\left( [(\alpha ,\alpha )]_{q}\right)^{-l}
(ad\ {e'}_{\delta})^{l}e_{\b} \ ,
$$
$$
{e'}_{l\delta}=
[e_{\alpha+(l-1)\delta},e_{\b}]_{q}
$$
and, finally,
$$
(q-q^{-1})E(z)= \log \left( 1+(q-q^{-1}){E'}(z)\right)
$$
where $E(z)$ and ${E'}(z)$ are generating functions for
$e_{n\d}$ and for ${e'}_{n\d}$:
$$E(z)=\sum_{n \geq 1}e_{n\d}z^{-n},$$
$${E'}(z)=\sum_{n \geq 1}{e'}_{n\d}z^{-n}.$$
The negative root vectors are given by the rule
$e_{-\gamma}$ = $e_{\gamma}^{*}$.

The universal $R$-matrix for $U_q(\widehat{sl}_2)$ has the following form
 \cite{hor}:
\[
{\cal R}=\left(\prod_{n\geq 0}^{\rightarrow} \exp_{q_{\alpha}}
\left( (q-q^{-1})e_{\alpha+n\delta} \otimes e_{-\alpha-n\delta}
\right)\right)\cdot
\]
\[
\exp\left( \sum_{n>0}(q-q^{-1})\frac{n(e_{n\delta}
\otimes e_{-n\delta})}{[n(\a ,\a)]_q}\right)\cdot
\]
\bn
\left(\prod_{n\geq 0}^{\leftarrow}\exp_{q_{\alpha}}\left( (q-q^{-1})
e_{\beta+n\delta}\otimes e_{-\beta-n\delta}\right)\right)\cdot {\cal K}
,
\label{urm}
\ed 
where the order on n is direct in the first product and it is inverse
in the second one. Factor ${\cal K}$ is defined by the  formula:
$$
{\cal K}=q^{{\frac{h_\a\otimes h_\a}{(\a ,\a)}}}
$$
%%%%%%%%%%%%%%%%%%%%%%%%%%%%%%%%%%%%%%%%%%%%%%%%%%%%%%%%%%%%%%%%%%%5

\section {The Structure of $ V_+(\l)\otimes V_\half (\l)$ Square}
In this Appendix we prove Baxter eq. (\ref{Beq})
for Q-operators (\ref{rep1}-\ref{rep2}) defined in {\bf Section 3}.
For simplicity we consider the special case of representations
(\ref{rep1}-\ref{rep2})
\bn
e_{\a_1}=e\hsp e_{\a_0}=\frac{\l}{(q-q^{-1})^2} e^{-1}
\hsp k_{a_1}=K,
\label{genv}
\ed
where $Ke= q^2 eK$.
The representation 
(\ref{genv})
of $\bm$ algebra for $\RMAT$ describes Volterra model.

Introduce the representation  
(\ref{genv}) space spanned by $|j\>,\,\,\,j\in \Bbb{Z}$,
with the action of $e,\,K$ generators 
\bn
e\, |j\>=q^{-j}\,|j+1\>\hsp
K\, |j\>=q^{2j}\,|j\>
\label{commrel}
\ed

For describing the space  $ V_+(\l)\otimes V_\half (\l)$
one needs also the evaluation representation  $ V_\half (\l)$  (\ref{ev}) with generators
$$
e_{\a_1}=e_\a,\hsp
e_{\a_0}= \l e_{-\a},\hsp
k_{\a_1}=k.
$$
We have the action of $\s$ algebra 
on 2-dimensional $\s$ space  ($\,|+\>,\,\,\, \,|-\>$)
$$
e_{\a} \,|+\>=0\hsp e_{\a} \,|-\>= \,|+\>, 
$$
$$
e_{-\a} \,|+\>=\,|-\>\hsp e_{-\a} \,|-\>= \,|+\>, 
$$
$$
k\,|\pm\>=q^{\pm 1}\,|\pm\>.
$$
Algebra $\bp$ acts in square  $ V_+(\l)\otimes V_\half (\l)$
by comultiplication (\ref{comult})
$$
\Delta (k_{\alpha_{i}})=K \otimes k,
$$
\bn
\Delta (e_{\alpha_{1}}) = 
e\otimes k+
1\otimes e_{\alpha},
\label{evcomult}
\ed
$$
\Delta (e_{\alpha_{0}}) = 
\l(\frac{e^{-1}}{(q-q^{-1})^2}\otimes k^{-1}+ 1\otimes e_{-\alpha})
$$
Now we should check that in the square  $ V_+(\l)\otimes V_\half (\l)$
spanned by vectors
\bn
\,|j\>\otimes \,|\pm\>, \hsp j\in \Bbb{Z}
\label{vects}
\ed
there exists subrepresentation isomorphic to  $ V_+(\l q^{-2})$.

For this we choose linear combinations
$$
U_j= \a_j \,|j\>\otimes \,|+\> +\b_j \,|j+1\>\otimes \,|-\>, \hsp j\in \Bbb{Z}
$$
of vectors with the same grading
from the space (\ref{vects}),
and act on vectors $U_j$ with generators
$\Delta (e_{\alpha_{1}})$ and
$\Delta (e_{\alpha_{0}})$
(\ref{evcomult}).

For 
$$
\a_j= \frac{q^{-j+1}}{1-q^2}\hsp
\hbox{and}
\hsp
\b_j= q^{-2j}, \hsp j\in \Bbb{Z},
$$
one has
\bn
\Delta (e_{\alpha_{1}})\, U_j= q^{-j}\, U_{j+1},
\hsp
\Delta (e_{\alpha_{0}}) \,U_j= \frac{\l q^{-2}}{(q-q^{-1})^2}
q^{j-1} \,U_{j-1}.
\label{cogenv}
\ed

Comparing actions of generators (\ref{genv}-\ref{commrel}) and  (\ref{cogenv})
one concludes that there exists subrepresentation
$V_+(\l q^{-2})$
in the square  $ V_+(\l)\otimes V_\half (\l)$.

To take factor space  $ V_+(\l)\otimes V_\half (\l) /  V_+(\l q^{-2}) $
we act with  generators
$\Delta (e_{\alpha_{1}})$,
$
\Delta (e_{\alpha_{0}})
$
on vectors
$$
W_j= q^j  \,|j\>\otimes \,|+\>,\hsp j\in \Bbb{Z},\hsp (\hbox{modulo}\,\, U_j)
$$
and get
$$
\Delta (e_{\alpha_{1}})\, W_j= q^{-j}\, W_{j+1},
\hsp
\Delta (e_{\alpha_{0}})\, W_j= \frac{\l q^{2}}{(q-q^{-1})^2}\,\,
q^{j-1} \,W_{j-1}
\hsp (\hbox {modulo}\,\, U_j) .
$$
So we prove that the factor  $ V_+(\l)\otimes V_\half (\l) /  V_+(\l q^{-2}) $
is isomorphic to $ V_+(\l q^{2}) $.

The same procedure can be easily repeated for the general case of 
(\ref{rep1}-\ref{rep2}).

\end{document}